\documentstyle[prd,aps,psfig]{revtex}

\begin{document}
\draft
\twocolumn[\hsize\textwidth\columnwidth\hsize\csname
@twocolumnfalse\endcsname
\preprint{PACS: 98.80.Cq, IJS-TP-97/16,\\
{}~hep-ph/yymmnn}

\newcommand\lsim{\mathrel{\rlap{\lower4pt\hbox{\hskip1pt$\sim$}}
    \raise1pt\hbox{$<$}}}
\newcommand\gsim{\mathrel{\rlap{\lower4pt\hbox{\hskip1pt$\sim$}}
    \raise1pt\hbox{$>$}}}

\title{$b-\tau$ unification and large atmospheric mixing: 
a case for non-canonical see-saw}

\author{Borut Bajc$^{(1)}$, Goran Senjanovi\'c$^{(2)}$
and Francesco Vissani$^{(3)}$ }

\address{$^{(1)}${\it J. Stefan Institute, 1001 Ljubljana, Slovenia}}

\address{$^{(2)}${\it International Centre for Theoretical Physics,
34100 Trieste, Italy }}

\address{$^{(3)}${\it INFN, Laboratori Nazionali del Gran Sasso,
Theory Group, Italy}}

\date{\today}
\maketitle

\begin{abstract}
We study the second and third generation masses in the context of
the minimal renormalizable SO(10) theory. We show that if the
see-saw takes the non-canonical (type II) form, large atmospheric
neutrino mixing angle requires $b-\tau$ unification.
\end{abstract}


\vskip1pc]

{\it A. Introduction}. \hspace{0.5cm}
A suspected quark-lepton symmetry is, as we know, badly broken
by the difference in their mixing angles. Small $V_{CKM}$ mixing
should be contrasted with the maximal mixing for atmospheric
neutrinos and probably large mixing for solar neutrinos. Why is
this so? This has become one of the major issues in the so-called
fermion mass and mixing problem.

In this Letter we address this question in the minimal
renormalizable SO(10) theory, without any additional symmetries
or interactions. We focus only on the second and
third generations for three reasons:

(i) in this case the neutrino mixing angle is maximal and
experimentally established;

(ii) it is much likely that in the case of the first family
we can not ignore higher dimensional operators;

(iii) in this simple $2\times 2$ case we can actually present
analytic expressions.

Our main result is the following. We show that in the case of
non-canonical see-saw, large neutrino mixing angle requires
$b-\tau$ unification. The rest of the paper is a proof of
this statement and a discussion of its implications.

The choice of SO(10) theory is highly natural. It unifies a family
of fermions; it unifies their interactions (except for gravity); it
has a see-saw mechanism \cite{Mohapatra:1981yp} of small neutrino
mass naturally built in; it has charge conjugation as a gauge
symmetry; and, in its supersymmetric version, leads naturally to a
theory of R-parity \cite{Aulakh:1999cd,Aulakh:2000sn}.

The last result holds true in the renormalizable version of the
theory with a $126_H$ dimensional Higgs supermultiplet
used to give masses to the right-handed neutrinos.

\vspace{0.2cm}

{\it B. Canonical (type I) versus non-canonical (type II)
see-saw mechanism}. \hspace{0.5cm}
The minimal Higgs that breaks SU(2)$\times$U(1) symmetry and gives
mass to the fermions is under the Pati-Salam
SU(2)$_L\times$SU(2)$_R\times$SU(4)$_C$ symmetry

\begin{equation}
10_H=(2,2,1)+(1,1,6)\;,
\end{equation}

\noindent
and so $\langle 10_H\rangle =\langle (2,2,1)\rangle\ne 0$ implies
the well-known quark-lepton symmetric relation for fermion masses

\begin{equation}
m_D=m_E\;,
\end{equation}

\noindent
which works well for the $3^{rd}$ family, and fails badly for
the first two. You can correct this by adding more Higgses, or
appealing to higher dimensional operators (see for example
\cite{Albright:1998vf,Babu:1998wi,Blazek:2002ta}). However, a nice
and important point was raised around twenty years ago
\cite{Lazarides:1980nt}. Ten years ago Babu and Mohapatra utilized it
to study neutrino masses and mixings \cite{Babu:1992ia}. With $10_H$
and $126_H$ the Yukawa sector of the Lagrangian is given by

\begin{equation}
{\cal L}_Y=10_H\psi Y_{10}\psi+126_H\psi Y_{126}\psi\;,
\end{equation}

\noindent
where $\psi$ stands for the $16$ dimensional spinors which
incorporate a family of fermions, and $Y_{10}$ and $Y_{126}$
are the Yukawa coupling matrices in generation space.

From

\begin{equation}
126_H=(3,1,10)+(1,3,\overline{10})+(2,2,15)+(1,1,6)
\end{equation}

\noindent
one has

\begin{equation}
M_{\nu_R}=Y_{126}\langle (1,3,\overline{10})_{126}\rangle\;,
\end{equation}

\noindent
where $\langle(1,3,\overline{10})_{126}\rangle =M_R$, the scale of
SU(2)$_R$ gauge symmetry breaking.

It can be shown that, after the SU(2)$\times$U(1) breaking through
$\langle 10_H\rangle =\langle (2,2,1)\rangle\approx M_W$, the
$(3,1,10)$ multiplet from $126_H$ gets a small vev
\cite{Mohapatra:1980yp,Magg:1980ut}

\begin{equation}
\langle (3,1,10)_{126}\rangle\propto{M_W^2\over M_{parity}}\;,
\end{equation}

\noindent
where $M_{parity}$ is the scale of the breakdown of parity. In 
general $M_R$ and $M_{parity}$ are not necessarily equal, but 
typically one breaks parity through the breaking of SU(2)$_R$ 
symmetry, in which case $M_R=M_{parity}$. This is what we take 
hereafter. 

In turn, neutrinos pick up small masses

\begin{equation}
\label{seesaw}
M_{\nu_L}=Y_{126}\langle (3,1,10)_{126}\rangle +
m_D^TM_{\nu_R}^{-1}m_D\;,
\end{equation}

\noindent
where $m_D$ is the neutrino Dirac mass matrix.
It is often assumed, for no reason whatsoever, that the second term
dominates. This we call canonical (often called type I) see-saw.
In what follows we explore the opposite case, which
we call non-canonical (type II) see-saw.
After all, it does not involve Dirac mass terms and so there is no
reason a priori in this case to expect quark-lepton analogy of
mixing angles. In this sense the non-canonical see-saw is
physically more appealing.
More than that, we will show that the large leptonic
mixing fits perfectly with the small quark mixing, as long as
$m_b=m_\tau$.

The crucial ingredient is the fact \cite{Babu:1992ia} that through
a non-vanishing tadpole a $(2,2,15)$ field in $126_H$ also picks up
a vev:

\begin{equation}
\langle (2,2,15)_{126}\rangle\approx\left({M_R\over M_{GUT}}\right)^2
\langle (2,2,1)_{10}\rangle\;.
\end{equation}

In the supersymmetric version of the theory this requires a
$210$ dimensional Higgs at the GUT scale.

\vspace{0.2cm}

{\it C. Non-canonical see-saw: $b-\tau$ unification and large
atmospheric neutrino mixing.} \hspace{0.5cm}
Most of the study throughout the years assumed the canonical see-saw,
i.e. the second term dominates in (\ref{seesaw}). The original claim
of \cite{Babu:1992ia} that the leptonic mixing matrix $V_l$ had a
small $2-3$ element was questioned by using a non-minimal model
\cite{Oda:1998na} or the freedom to adjust the phases in the
mixing matrices \cite{Matsuda:2001bg}.
Last year we studied \cite{Bajc:2001fe} the
opposite case, the non-canonical see-saw and noticed that it fitted
nicely with a large $2-3$ mixing angle responsible for atmospheric
neutrinos.

We give here a simple argument in favour of this.
We show how maximal $\mu-\tau$ mixing fits nicely with $b-\tau$
unification.

To see this, notice that fermion masses take the following form

\begin{eqnarray}
\label{mu}
M_U&=&Y_{10}v_{10}^u+Y_{126}v_{126}^u\;,\\
M_D&=&Y_{10}v_{10}^d+Y_{126}v_{126}^d\;,\\
M_E&=&Y_{10}v_{10}^d-3Y_{126}v_{126}^d\;,\\
\label{mn}
M_N&=&Y_{126}\langle (3,1,10)_{126}\rangle\;,
\end{eqnarray}

\noindent
where $U$, $D$, $E$, $N$ stand for up quark, down quark,
charged lepton and neutrino, respectively, while
$v_{10}^{u,d}$ and $v_{126}^{u,d}$ are the two
vevs of $(2,2,1)$ in $10_H$ and $(2,2,15)$ in $126_H$,
and the last formula is the assumption of the
non-canonical see-saw. The result is surprisingly simple.
Notice that \cite{Brahmachari:1997cq}

\begin{equation}
\label{mnde}
M_N\propto Y_{126}\propto M_D-M_E\;.
\end{equation}

Now, let us study the $2^{nd}$ and $3^{rd}$ generations,
and work in the basis of $M_E$ diagonal. The puzzle then is:
why a small mixing in $M_D$ corresponds to a large mixing in
$M_N$? For simplicity take the mixing in $M_D$
to vanish, $\theta_D=0$, and ignore the second generation masses,
i.e. take $m_s=m_\mu=0$. Then

\begin{eqnarray}
\label{mnu}
M_N\propto\pmatrix{
  0
& 0
\cr
  0
& m_b-m_\tau
\cr}\;.
\end{eqnarray}

Obviously, unless $m_b=m_\tau$, neutrino mixing vanishes. Thus,
large mixing in $M_N$ (the physical leptonic mixing in the
above basis) is deeply connected with the $b-\tau$
unification. Notice that we have done no model building whatsoever;
we only assumed a renormalizable SO(10) theory and the non-canonical
see-saw.

Before we discuss (\ref{mnu}) more carefully by switching on $m_\mu$,
$m_s$ and the mixings, let us comment on the implication of our
result. First, notice that it does not depend on the number of
$10_H$'s. Notice also that it is not easily generalized to three
generations, i.e. it is not easy to give the same reasoning why
the solar neutrino mixing should be large (to be confirmed
experimentally).

In short, our results should be taken as an argument in favour of
the non-canonical see-saw: large atmospheric
mixing angles and $b-\tau$ unification seem to prefer clearly
this form of the see-saw mechanism.

\vspace{0.2cm}

{\it D. Quantitative analysis.} \hspace{0.5cm}
Let us now be more quantitative and turn on $m_s$, $m_\mu$ and
$\theta_D$. Notice that $\theta_D$ is not a $2-3$ $V_{CKM}$
mixing angle, but rather a difference between charged lepton
and down quark mixing angles (recall that we choose $M_E$ diagonal).

An important comment. It is not that all the $32$ doublets in
$(2,2,1)$ and $(2,2,15)$ remain light; with the minimal fine-tuning
we end up with only two of them at $M_Z$. Let us denote their vevs
by $v_i$ ($i=u,d$), where $M_W=g\sqrt{v_u^2+v_d^2}/2$ and we adopt
as usual $\tan{\beta}=v_u/v_d$. Then we can write

\begin{equation}
v_{10}^i=v_i\cos{\alpha_i}\;\;\;,\;\;\;
v_{126}^i=v_i\sin{\alpha_i}\;\;\;,\;\;\;
(i=u,d)\;\;\;,
\end{equation}

\noindent
where $\alpha_i$ are unknown angles. Defining

\begin{equation}
\label{defxy}
x={\tan{\alpha_u}\over\tan{\alpha_d}}\;\;\;,\;\;\;
y={\cos{\alpha_d}\over\cos{\alpha_u}}
\end{equation}

\noindent
(notice that either $x^2\le y^2\le 1$ or $x^2\ge y^2\ge 1$),
it is a simple exercise to derive from (\ref{mu})-(\ref{mn})

\begin{eqnarray}
\label{ye}
Y_E&=&{1\over 1-x}\left[4yY_U-(3+x)Y_D\right]\;,\\
\label{yn}
Y_N&=&c\left(Y_E-Y_D\right)\;,
\end{eqnarray}

\noindent
where $M_U=v_uY_U$, $M_D=v_dY_D$, $M_E=v_dY_E$, $M_N\propto Y_N$,
and $c$ is an unknown constant in this theory.
Since $Y$'s are symmetric, we can write for species $X$

\begin {equation}
Y_X=XY_X^dX^T\;,
\end{equation}

\noindent
where $Y_X^d$ are diagonal Yukawa matrices and $X$ are in general
unitary. In what follows we do not wish to play with the adjustment
of phases and so take $X$ to be orthogonal matrices for simplicity
and transparency.

Let $\theta_l$, $\theta_D$ and $\theta_q$ denote the rotation angles
in $E^TN$, $D^TE$ and $D^TU$ respectively ($\theta_l$ and $\theta_q$
are the leptonic and quark weak mixing angles respectively). From
(\ref{yn}) we get

\begin{equation}
\label{thl}
\tan{2\theta_l}={\sin{2\theta_D}\over
{y_\tau-y_\mu\over y_b-y_s}
-\cos{2\theta_D}}\;.
\end{equation}

Next, we wish to connect $\theta_D$ with $\theta_q$ in order to
have the dependence of $\theta_l$ with $\theta_q$.
>From (\ref{ye}) one has ($c_D=\cos{\theta_D}$,
$c_q=\cos{\theta_q}$, etc.)

\begin{eqnarray}
\label{xy}
\pmatrix{
   c_D^2y_\tau+s_D^2y_\mu-y_b
&  c_q^2y_t+s_q^2y_c
\cr
   s_D^2y_\tau+c_D^2y_\mu-y_s
&  s_q^2y_t+c_q^2y_c
\cr}
\pmatrix{
   x
\cr
   4y
\cr}=\nonumber\\
\pmatrix{
   c_D^2y_\tau+s_D^2y_\mu+3y_b
\cr
   s_D^2y_\tau+c_D^2y_\mu+3y_s
\cr}&\;.
\end{eqnarray}

After introducing

\begin{equation}
\label{epsilon}
\epsilon_u={y_c\over y_t}\;,\;
\epsilon_d={y_s\over y_b}\;,\;
\epsilon_e={y_\mu\over y_\tau}\;,\;
\epsilon={y_b-y_\tau\over y_b}\;,
\end{equation}

\noindent
and after some computational tedium we get from (\ref{ye})
and (\ref{xy})

\begin{eqnarray}
\label{alpha}
&&(1-\epsilon_e)\tan{\theta_D}
\left[(1-\epsilon_u\epsilon_d)\tan^2{\theta_q}+
(\epsilon_u-\epsilon_d)\right]=\nonumber\\
&&(1-\epsilon_u)\tan{\theta_q}
\left[(1-\epsilon_e\epsilon_d)\tan^2{\theta_D}+
(\epsilon_e-\epsilon_d)\right]\;.
\end{eqnarray}

In the limit $\epsilon_i=0$ ($i=u,d,e$) there are two solutions:
$\tan{\theta_D}=0$ and $\tan{\theta_D}=\tan{\theta_q}$.
The first solution can be shown to be unrealistic, whereas the
second one gives the important relation between the physical
mixing angles of quarks and leptons:

\begin{equation}
\label{thl0}
\tan{2\theta_l}={\sin{2\theta_q}\over 2\sin^2{\theta_q}-\epsilon}\;.
\end{equation}

Since $\theta_q=\theta_{bc}$ of $V_{CKM}$, $\theta_q\approx 10^{-2}$,
(\ref{thl0}) shows manifestly that $\tan{\theta_l}\approx 1$
requires $\epsilon\approx 0$, i.e. $y_b\approx y_\tau$ as we
argued repeatedly.

Let us now switch on the second generation masses, i.e. let us
take $\epsilon_i\ne 0$. From (\ref{alpha}) one can see that
the physically acceptable solution is

\begin{equation}
\tan{\theta_D}={\cal O}(\delta)\;\;\;,\;\;\;
\delta=\epsilon_i,\tan{\theta_q}\approx 10^{-2}\;\;\;.
\end{equation}

From (\ref{thl}) is then obvious that $b-\tau$ unification
$y_\tau=y_b+{\cal O}(\delta)$ is sufficient to make the mixing angle
large, i.e. $\tan{2\theta_l}={\cal O}(1)\gg\delta$. This is our main
result, rather nontrivial in our opinion. A small quark mixing angle
automatically leads to a large leptonic mixing in the $2-3$ case.

\vspace{0.2cm}

{\it E. From high to low energy: running.} \hspace{0.5cm}

Our expressions are valid at the unification scale $M_{GUT}$.
Thus we must run the physical parameters from $M_{GUT}$ to
$M_Z$ in order to be precise. However, in this case the running
is not so important as it may seem. Namely, in this letter we
want to study the implications of the
SO(10) symmetry (in its minimal renormalizable version) on
fermion masses and mixings. What we said up to now is equally valid
in ordinary and supersymmetric (with $210_H$ Higgs) SO(10) gauge theory.
We wish to emphasize the generic feature of the model, that is the
connection between the large $\theta_{atm}$ and $b-\tau$ unification
and do not worry so much about the precise numerical estimates.
This requires specifying precisely the nature of the low energy
effective theory. Still, it is instructive to see the impact of
running. We thus discuss briefly the supersymmetric case and
leave the complete discussion for a longer paper now in preparation.

The neutrino matrix elements $M_{ij}$ run at the 1-loop level and
neglecting threshold effects according to
\cite{Chankowski:1993tx,Babu:qv,Balaji:2000gd}

\begin{equation}
16\pi^2{d\over dt}M_{ij}=
\left[y_\tau^2(k_i+k_j)+6y_t^2-6g_2^2-{6\over 5}g_1^2\right]
M_{ij}\;,
\end{equation}

\noindent
where $t=\ln{(Q/M_Z)}$, $g_1$ is normalized in the SU(5) fashion,
$i,j=2,3$ stand for the second and third generations, and
$k_2=0$, $k_3=1$. The neutrino mixing angle at
the electroweak scale is

\begin{eqnarray}
\tan{2\theta_l}|_{M_Z}&=&{2M_{23}(0)\over M_{22}(0)-M_{33}(0)}\nonumber\\
&=&{2M_{23}(t_{GUT})B_\tau\over
M_{22}(t_{GUT})-B_\tau^2M_{33}(t_{GUT})}\;,
\end{eqnarray}

\noindent
where $t_{GUT}=\ln{(M_{GUT}/M_Z)}$ and

\begin{equation}
B_\tau=\exp{\left(-{1\over 16\pi^2}
\int_0^{t_{GUT}}y_\tau^2(t)dt\right)}\;.
\end{equation}

The elements $M_{ij}(t_{GUT})$ are exactly the ones discussed
throughout the paper. We can thus recalculate (\ref{thl}) (valid
at $M_{GUT}$) at $M_Z$:

\begin{equation}
\tan{2\theta_l}|_{M_Z}={B_\tau\sin{2\theta_D}\over
{y_\tau-y_\mu\over y_b-y_s}
-{1+B_\tau^2\over 2}\cos{2\theta_D}
-{1-B_\tau^2\over 2}\left(1+2{y_\tau-y_b\over y_b-y_s}\right)
}\;.
\end{equation}

%\begin{equation}
%\tan{2\theta_l}|_{M_Z}={B_\tau\sin{2\theta_D}\over
%{\epsilon_d-\epsilon_e+\epsilon\epsilon_e
%-B_\tau^2\epsilon\over 1-\epsilon_d}
%+(1+B_\tau^2)\sin^2{\theta_D}}\;.
%\end{equation}

All the parameters of the right-hand-side are to be evaluated at the
GUT scale. For this reason the same equation (\ref{alpha})
is again used to express $\theta_D$. Clearly, as before, large
neutrino mixing angle comes out as soon as $y_b$ and $y_\tau$
unify at the GUT scale. Of course, the precise value of the neutrino
mixing angle depends on this running, however the qualitative
behaviour does not change.

A more detailed approach would require to use numerical techniques
to account for (1) the running as function of $\tan{\beta}$; (2)
the inclusion of threshold corrections \cite{Hall:1993gn}; and
(3) first generation effects. However, threshold effects in SO(10)
are bound to be important and high precision calculations may
actually not be so useful, see for example \cite{Dixit:1989ff}.

What about the values of neutrino masses? We do not enter into this
issue here since we have no new results beyond \cite{Bajc:2001fe}.

\vspace{0.2cm}

{\it F. Summary and outlook.} \hspace{0.5cm}

The sharp contrast of quark and lepton mixings is often considered a
deep puzzle. We argued here that it is actually quite natural in
the minimal SO(10) renormalizable theory. All that is required is
that the see-saw mechanism takes a non-canonical form free from
Dirac masses. The approximate formula (\ref{thl0}) expresses it
clearly: a small quark mixing $\theta_q=\theta_{bc}\approx 10^{-2}$
gives naturally a large $\theta_l=\theta_{atm}$ if $\epsilon\approx 0$,
i.e. $y_b\approx y_\tau$. Actually, the essence of our work lies in
formulae (\ref{mnde})-(\ref{mnu}). Formula (\ref{mnu}), valid in the
approximation of vanishing second generation masses and vanishing
quark and lepton mixings speaks eloquently: unless $m_b=m_\tau$ at
the large scale, we will have a vanishing atmospheric neutrino mixing.
In short, the non-canonical see-saw marries nicely $b-\tau$ unification
with the maximal atmospheric neutrino mixing. This can be of great
help in trying to pin-point the nature of the see-saw mechanism: our
study points in favor of the non-canonical version.

Strictly speaking, a numerical study showed that in the $3\times 3$
case, by playing with CP phases, even the canonical see-saw can be
made to work \cite{Matsuda:2001bg}. However, in our case,
the $2-3$ family study offers
physical insight into the question, and after all the first family of
fermions may suffer from the higher dimensional operators. The $10_H$
and $126_H$, the minimal Higgses needed to give masses to all fermions,
work beautifully: $10_H$ offers $m_b=m_\tau$, and $126_H$ offers
$3m_s=-m_\mu$ at the GUT scale; and in this framework
a small $\theta_{cb}$ ($\theta_{ts}$) and a large $\theta_{atm}$
become naturally connected. Thus, the observational evidence that
quarks and leptons have sharply different mixing angles fits nicely
with the belief that they are one and the same object at a
fundamental level.

\vskip 0.3cm

We are grateful to Rabi Mohapatra for his encouragement, and to
Alejandra Melfo for useful comments and a careful reading of the
manuscript.
The work of B.B. is supported by the Ministry of Education, Science
and Sport of the Republic of Slovenia.
The work of G.S. is partially supported by EEC, under
the TMR contracts ERBFMRX-CT960090 and HPRN-CT-2000-00152.
We express our gratitude to INFN, which
permitted the development of the present study
by supporting an exchange program with the
International Centre for Theoretical Physics.

\end{document}